\documentclass[10pt,conference]{IEEEtran}
\IEEEoverridecommandlockouts

\usepackage{cite}
\usepackage{amsmath,amssymb,amsfonts}
\usepackage{algorithm2e}
\usepackage{graphicx}
\usepackage{textcomp}

\def\BibTeX{{\rm B\kern-.05em{\sc i\kern-.025em b}\kern-.08em
    T\kern-.1667em\lower.7ex\hbox{E}\kern-.125emX}}

\usepackage{packages}

\begin{document}

\title{Data Preparation for Fairness-Performance Trade-Offs: A Practitioner-Friendly Alternative?}

\author{\IEEEauthorblockN{Gianmario Voria} 
\IEEEauthorblockA{University of Salerno\\
Salerno, Italy\\
gvoria@unisa.it}
\and
\IEEEauthorblockN{Rebecca Di Matteo}
\IEEEauthorblockA{University of Salerno\\
Salerno, Italy\\
r.dimatteo10@studenti.unisa.it}
\and
\IEEEauthorblockN{Giammaria Giordano}
\IEEEauthorblockA{University of Salerno\\
Salerno, Italy\\
giagiordano@unisa.it}
\and
\IEEEauthorblockN{Gemma Catolino}
\IEEEauthorblockA{University of Salerno\\
Salerno, Italy\\
gcatolino@unisa.it}
\and
\IEEEauthorblockN{Fabio Palomba}
\IEEEauthorblockA{University of Salerno\\
Salerno, Italy\\
fpalomba@unisa.it}}

\maketitle

\begin{abstract}

\textit{Context.} As machine learning (ML) systems are increasingly adopted across industries, addressing fairness and bias has become essential. While many solutions focus on ethical challenges in ML, recent studies highlight that data itself is a major source of bias. Pre-processing techniques, which mitigate bias before training, are effective but may impact model performance and pose integration difficulties. In contrast, fairness-aware \emph{`Data Preparation'} practices are both familiar to practitioners and easier to implement, providing a more accessible approach to reducing bias.
\textit{Objective.} This registered report proposes an empirical evaluation of how optimally selected fairness-aware practices, applied in early ML lifecycle stages, can enhance both fairness and performance, potentially outperforming standard pre-processing bias mitigation methods. \textit{Method.} To this end, we will introduce \textsc{FATE}, an optimization technique for selecting \emph{`Data Preparation'} pipelines that optimize fairness and performance. Using \textsc{FATE}, we will analyze the fairness-performance trade-off, comparing pipelines selected by \textsc{FATE} with results by pre-processing bias mitigation techniques.

\end{abstract}

\begin{IEEEkeywords}
Machine Learning Fairness; Data Preparation; Genetic Algorithm; Empirical Software Engineering.
\end{IEEEkeywords}

\section{Introduction}
The rapid adoption of machine learning (ML) has raised ethical concerns around \emph{fairness} \cite{mehrabi2021survey}---the principle that models should make impartial decisions without introducing biases for or against specific groups. Unfairness arises when models replicate biases from training data \cite{pagano2023bias,pessach2022review}, leading to decisions that create ethical and legal risks \cite{miller2019machine}.

To address fairness, the software engineering (SE) research community—especially within software engineering for artificial intelligence (SE4AI)—has advanced bias mitigation techniques \cite{hort2024bias}. These approaches generally fall into three categories: \emph{pre-processing}, \emph{in-processing}, and \emph{post-processing} techniques. Pre-processing methods, like \textsc{FairSMOTE} \cite{chakraborty2021bias}, work to mitigate bias by rebalancing sensitive groups within the dataset prior to training. In-processing techniques adjust the learning algorithm itself to minimize bias, such as fairness-aware training methods \cite{li2022training}. Post-processing techniques aim to improve fairness by adjusting model outputs, for example, through fairness testing \cite{galhotra2017fairness}.

A key focus in fairness research is the recognition that bias in ML systems often stems from imbalanced datasets, making early-stage interventions in the \emph{`Data Preparation'} phase, i.e., the ML engineering stage in which raw data gets processed into a clean, structured, and optimized format suitable for efficient model training \cite{burkov2020machine}, critical for achieving fair outcomes \cite{Biswas2021981, Valentim2019391}. While specialized data pre-processing techniques \cite{chakraborty2021bias,kamiran2012data,feldman2015certifying}, i.e., specific techniques and algorithms designed to reduce ML models' bias operating on the dataset before model training, have shown effectiveness in mitigating bias, they also come with considerable costs. These costs include both the additional computation demands they place on model development and the extra effort required for practitioners to apply them as they diverge from standard workflows \cite{toolkit_survey, toolkit_landscape}. In contrast, recent research highlighted the potential efficacy in mitigating bias of common \emph{`Data Preparation'} practices, i.e., ML engineering techniques used to increase models' efficacy and that operate before model training \cite{voria2024mapping}. Examples of these practices are data scaling, resampling, or normalization \cite{voria2024mapping}, and they differ from specialized bias mitigation pre-processing techniques because they are not explicitly built to mitigate bias. These techniques, named \textit{fairness-aware practices}, are lightweight, often familiar to practitioners, and seamlessly integrated into ML workflows, making them inherently more accessible and practical to implement \cite{voria2024survey}. In addition, they differ from general pre-processing strategies because they are tailored to specific goals or domain requirements.

Previous literature suggests that \textit{fairness-aware practices} are not only more practical but also often preferred by practitioners over \textit{specialized bias mitigation methods} for addressing fairness issues. For instance, Voria et al. \cite{voria2024mapping} documented that fairness-aware practices primarily cluster around the \emph{`Data Preparation'} phase. At the same time, a survey of expert ML practitioners confirmed that these practices positively influence fairness and are frequently used to modify datasets relative to protected attributes \cite{voria2024survey}. Additionally, these practices appear to offer a distinct advantage in managing the fairness-performance trade-off, as they can enhance both fairness and model performance without the potential performance costs sometimes associated with bias mitigation techniques \cite{Raina2022}.
These insights lead us to hypothesize that there may be a practical advantage in using data preparation practices for fairness, which practitioners are already familiar with and regularly employ. Accordingly, this study aims to empirically examine this potential advantage by investigating whether and to what extent standard data preparation practices can effectively mitigate bias. Our confirmatory study will also assess how these approaches compare against specialized bias mitigation algorithms, providing insights into their relative efficacy on the fairness-performance trade-off.

\steResearchQuestionBox{\faLightbulbO \hspace{0.05cm} \textbf{Working Hypothesis.} Based on the considerations that (1) specific pre-processing bias mitigation techniques are harder to apply than common fairness-aware practices, (2) fairness-aware practices at the \emph{`Data Preparation'} stage are known to have a positive impact on fairness and are also used to improve models' efficacy, and (3) practitioners already actively employ fairness-aware practices in their ML pipelines, our research hypothesis is that a near-optimal selection of fairness-aware practices during the \emph{`Data Preparation'} stage has a high positive impact on fairness and performance in practice and can be beneficial for practitioners to easily mitigate ML bias.}

Building on this hypothesis, the \textbf{objective} of this registered report is to outline a \textit{confirmatory} study that will empirically assess how a near-optimal selection of fairness-aware practices in earlier ML lifecycle stages can contribute to fair and efficient ML models. As pre-processing techniques represent state-of-the-art bias mitigation at these stages, our hypotheses will be structured around a comparison of near-optimal \emph{`Data Preparation'} pipelines against such solutions in terms of fairness and performance gains. Our contributions are threefold. First, we will develop an optimization technique, \textsc{FATE}, which selects \emph{`Data Preparation'} pipelines to optimize both fairness and performance. Second, we will assess how \textsc{FATE} supports balancing the fairness-performance trade-off in model building. Lastly, we will compare the impact of \textsc{FATE}-selected pipelines with widely used pre-processing bias mitigation techniques.

\section{Background \& Related Work}

ML fairness, i.e., the absence of prejudice toward a protected group by an automated decision maker \cite{mehrabi2021survey}, has emerged as a pivotal research domain within the \textit{Software Engineering} community, with an expanding body of literature examining the topic from various perspectives \cite{pessach2022review, starke2021fairness, chen2024fairness, hort2024bias}. This concern is highlighted by several known ethical incidents involving ML applications\cite{brun2018software,ia_ethical_incidents}, underscoring the urgent need for fair ML practices. Researchers have explored fairness through empirical studies, proposing bias mitigation strategies at different stages of the ML pipeline, named pre-, in-, and post-processing techniques. \textbf{Pre-processing} methods tackle bias in the training data. For example, Chakraborty et al. \cite{chakraborty2021bias} introduced \textsc{Fair-SMOTE}, a synthetic augmentation technique that preserves model performance. Reweighting methods, like those by Kamiran and Calders \cite{kamiran2012data}, adjust instance weights to improve fairness. \textbf{In-processing} techniques modify the learning algorithm itself to mitigate bias during training. Zhang et al. \cite{zhang2018mitigating} used an adversarial approach, while Chakraborty et al. \cite{chakraborty2020fairway} emphasized balancing fairness and performance through multi-objective optimization. Finally, \textbf{post-processing} techniques adjust model outputs after training to ensure fairness. Galhotra et al. \cite{galhotra2017fairness} proposed \textsc{Themis}, a tool that identifies bias through input perturbation, and Udeshi et al. \cite{udeshi2018automated} developed \textsc{Aequitas}, which enhances bias detection efficiency. Aggarwal et al. \cite{aggarwal2019black} proposed a black-box fairness testing method, while Zhang et al. \cite{zhang2020white} introduced a white-box approach using adversarial sampling to detect and mitigate biases. The research community and organizations have developed instruments to make these solutions available for software practitioners, referred to as fairness toolkits, comprising ready-to-use metrics to measure fairness or bias mitigation techniques \cite{toolkit_survey}.

\subsection{Research Objective and Motivation}
This registered report explores the hypothesis that applying near-optimal fairness-aware practices during the \emph{`Data Preparation'} stage could effectively bring to positive gains in both performance and fairness. Hence, our \textbf{final objective} is to perform an empirical study to test this hypothesis. To enable this study, we will develop a fairness-optimization technique specifically tailored to the \emph{`Data Preparation'} process that selects near-optimal pipelines of fairness-aware practices at this stage. This aligns with previous research: Gonzalez et al. \cite{gonzalez2023preprocessing}, starting from a similar hypothesis, developed \textsc{FairPipes}, a framework that optimizes the selection of ML pipelines, evaluating user-defined combinations of fairness and accuracy. However, our study will differ for three main reasons: (1) we will perform an empirical evaluation of our solution with a specific focus on its impact on the \textit{fairness-accuracy trade-off}; (2) our optimization technique is built on recent \emph{Software Engineering} research on fairness-aware practices \cite{voria2024survey} rather than general techniques, hence allowing us to expect a higher positive impact on fairness; (3) we will ground our work on recent \emph{Software Engineering} research on fairness metrics \cite{majumder2023fair} to optimize pipelines on the most suitable set of measures.

\section{FATE: Fairness-Aware Trade-Off Enhancement}
To enable our study and assess the extent to which a near-optimal set of fairness-aware practices can stand as a valuable support to mitigate bias in earlier stages of ML development, we will develop a meta-approach named \textsc{FATE} (\underline{F}airness-\underline{A}ware \underline{T}rade-Off \underline{E}nhancement). It will be a novel genetic algorithm (GA)-based \cite{whitley1994genetic} solution that aims to select near-optimal \emph{`Data Preparation'} pipelines through a linear combination of fairness and performance metrics, merging fairness and accuracy into a single, unified objective. \textsc{FATE} will be built aiming for two main objectives. First, we aim for a universally applicable solution, i.e., a solution that may optimize arbitrary \emph{`Data Preparation'} fairness-aware practices that are built to operate on a protected attribute in a dataset and which is, therefore, agnostic to the specific features the ML models enable and the domain it is applied to. Second, we aim for a solution that may be combined with additional bias mitigation algorithms, being therefore flexible enough to be used together with additional optimization solutions. As input, \textsc{FATE} will require the user to specify the optimization task, i.e., the ML model that will be used for evaluation, the dataset to be optimized, and its corresponding protected attribute necessary to perform a bias evaluation. This approach ensures \textsc{FATE} can generalize fairness-aware practices to any dataset and task, tailoring results to specific contexts.

The structure of \textsc{FATE} will be as shown in Algorithm \ref{alg1}. It simulates the process of natural selection by evolving a population of candidate solutions through selection, crossover, and mutation \cite{whitley1994genetic}. These steps are described below, alongside a \textit{running example} that follows each step.

\begin{algorithm}
\caption{Pseudocode of \textsc{FATE}.}
\label{alg1}
\SetAlgoLined
\KwIn{Optimization Task $O$, Dataset $D$, Protected Attribute $Attr$, Number of generations $G$, Population size $N$, Crossover Rate $\alpha$, Mutation Rate $\beta$}
\KwOut{Optimized data preparation pipeline}

\smallskip

$P \leftarrow \texttt{initializePopulation}(N)$\; 

\For{$g = 1$ \KwTo $G$}{
    
    \For{each individual $i \in P$}{
        $i.\texttt{fitness} \leftarrow \texttt{evaluateFitness}(i, D, Attr, O)$\;
    }

    $matingPool \leftarrow \texttt{bestIndividuals}(P)$\;
    
    \For{each pair of parents $(p_1, p_2) \in \texttt{matingPool}$}{
        $offsprings \leftarrow \texttt{crossover}(p_1, p_2, \alpha)$\;
        
        $offsprings \leftarrow \texttt{mutate}(offsprings, \beta)$\;
        
        $newPopulation.\texttt{add}(offsprings)$\;
    }
    
    $P \leftarrow newPopulation$\;
}
\Return $\texttt{bestIndividuals}(P)$\;
\end{algorithm}

\textsc{FATE} can be initially configured by four parameters \cite{hassanat2019choosing}: (1) the number of generations, determining how many iterations the algorithm undergoes to refine its population, (2) the population size, which sets the number of candidate solutions present in each generation, (3) the crossover rate, which indicates the probability that two chromosomes switch some of their genes, and (4) the mutation rate, which determines the chances of one gene to be randomly mutated.
\steResearchQuestionBox{\textbf{Running Example \#0.} The genetic algorithm is executed with a number of generations G = 10, a population size N = 5, a crossover rate $\alpha$ of the 25\%, and a mutation rate $\beta$ of the 25\% for introducing genetic diversity. The user specifies the following inputs: (1) binary classification using Logistic Regression (LR) as an optimization task, (2) a dataset that contains information on individuals with attributes such as age, gender, and income, and (3) \emph{gender} as the protected attribute to be evaluated for fairness.}

\smallskip \textbf{Step \#1. Individual Representation and Population.} In our context, each candidate solution represents a potential pipeline of \emph{`Data Preparation'} techniques: \textsc{FATE} will be designed to work with any practice belonging to this stage of the ML development, as long as they are implemented to optimize a certain aspect of a dataset relatively to a \emph{protected attribute}. For instance, a \textit{resampling technique} could modify class representation through incrementing samples belonging to protected groups, reducing bias \cite{Iosifidis20191375, GonzalezZelaya20192086}. Hence, the set of fairness-aware practices that an individual can select may vary as the user defines and implements specific techniques. In practice, individuals are defined by an assortment of fairness-aware practices that are randomly selected and applied in different combinations.

\steResearchQuestionBox{\textbf{Running Example \#1.} Five individuals get instantiated, each with a randomly selected set of fairness-aware \emph{`Data Preparation'} practices. 
\textsc{FATE} instantiates five pipelines, namely A: (Standard Scaling, Oversampling), B: (MinMax Scaling, Clustering), C: (Inverse Probability Weighting, Matching), D: (Resampling, Standard Scaling), E: (MinMax Scaling, Oversampling, Matching). After this step, the algorithm starts looping for ten generations before finding a near-optimal solution.}

\begin{table}
\footnotesize
	\centering
	\caption{Metrics used for the fitness evaluation.}
	\label{tab:metrics}
	\begin{tabular}{|p{3cm}|p{5cm}|}
		\hline
		\rowcolor{purple}
		\textcolor{white}{\textbf{Name}} & \textcolor{white}{\textbf{Operationalization}} \\ \hline
		 Precision (P)  & \((TP)/(TP + FP)\) \\
		\hline
	   Recall (R) & \((TP) /(TP + FN)\) \\
		\hline
            PR-AUC & \(\sum_{i=1}^{n-1} (R_{i+1} - R_i) \cdot P_{i+1}\) \\ \hline
        \rowcolor{purple!15}
  	Performance Score (PS) & \( \text{PR-AUC}  \) \\\hline

		Statistical Parity Difference (SPD)  & \( P(Y = 1|A = 0) - P(Y = 1|A = 1) \) \\\hline
		 Equal Opportunity Difference (EOD)  & \(TPR_{A=\text{unprivileged}} - TPR_{A=\text{privileged}} \)\\\hline
		Disparate Impact  (DI) &    \( P(Y = 1|A = 0))/(P(Y = 1|A = 1) \)  \\\hline
           \rowcolor{purple!15}
  	Fairness Score (FS) & \(|SPD| + |EOD| + |DI|\) \\
        \hline
        \hline
        \rowcolor{purple!15} \textbf{Fitness Function} & 
          \(\alpha \cdot PS - \beta \cdot FS\) \\\hline
	\end{tabular}
\end{table}

\smallskip \textbf{Step \#2. Fitness Evaluation and Selection.} Each individual is evaluated using a fitness function that measures how well it performs against the defined objectives. To perform this step, our algorithm will apply the \emph{`Data Preparation'} practices to the dataset provided as input, and afterward, it will train the ML algorithm specified as input on the prepared dataset to compute the metrics, using K-fold cross-validation to ensure that the whole dataset is used during the evaluation.

The fitness function will be designed to balance both predictive performance and fairness. For evaluating \emph{predictive performance}, we will rely on the \emph{Area Under Precision and Recall Curve (PR-AUC)} \cite{boyd2013aucpr}. This metric summarizes under a single measure information about Precision and Recall\cite{ml_metrics}, and has been shown to have a higher informative value compared to other metrics when evaluating binary classification tasks on imbalanced datasets \cite{saito2015precision}. \textit{PR-AUC} is computed using the area under the Precision-Recall (PR) curve by summing the contributions of individual segments, as shown in the formula in Table \ref{tab:metrics}. For each segment, the difference in recall values, \((R_{i+1} - R_i)\), represents the width, while the precision at the higher recall point, \(P_{i+1}\), approximates the height. To assess \emph{fairness}, we will use three key fairness metrics: \textsl{statistical parity difference} (SPD) \cite{agarwal2018reductions}, that compares the probability of favorable outcomes across groups, \textsl{equal opportunity difference} (EOD) \cite{hardt2016equality}, that assesses the difference in true positive rates between protected and unprotected groups, and \textsl{disparate impact ratio} (DI) \cite{feldman2015certifying}, that measures the ratio of favorable outcomes between groups. These metrics are widely accepted as standards in the evaluation of fairness in ML models \cite{franklin2022ontology,ferrara2024fairness,pagano2023bias,majumder2023fair}. Using these metrics, we evaluate how well the models treat different demographic groups equitably.

Since these two objectives—performance and fairness—often conflict \cite{mehrabi2021survey}, we will implement a multi-objective fitness function that balances this trade-off. For performance, a higher PR-AUC value, closer to 1, indicates a model with strong performance, achieving high precision and recall simultaneously. In contrast, a lower PR-AUC suggests that the model struggles to maintain precision as recall increases. Therefore, the \emph{performance score (PS)} will be equal to the \textit{PR-AUC} metric. Fairness metrics typically range from -1 to 1, with 0 representing the ideal unbiased value. To compute the \emph{fairness score (FS)}, we will minimize deviations from 0 by using the absolute values of each fairness metric. Finally, the fitness function will combine the performance score and fairness deviation, aiming to maximize predictive performance while minimizing fairness deviation. The final calculation will scale these two values using scaling factors, \(\alpha \text{ and } \beta\), which control the trade-off between performance and fairness. These factors can be adjusted based on the specific needs of the application, but they are set to the same equal value in our analysis to achieve balance among the measures, i.e., 0.5 each. This ensures that \textsc{FATE} prioritizes solutions with strong predictive performance and minimal fairness deviation.

Table \ref{tab:metrics} summarizes all the metrics used to calculate the fitness value and shows how they are computed. After this evaluation, the selection process will be guided by the best-performing individuals in terms of overall fitness value that will reproduce and form the next generation.

\steResearchQuestionBox{\textbf{Running Example \#2.} The individuals in the initial population get evaluated through the fitness function. Each pipeline is applied to the dataset, and the LR model selected as the optimization task is trained and evaluated. For instance, pipeline A achieves PS = 0.88 and FS = 0.10 (low fairness deviation). Pipeline C achieves PS = 0.90 and FS = 0.45 (higher fairness deviation). As the fitness function aims to balance these measures, and pipeline A scores better than pipeline C due to better fairness without highly deteriorating performances, pipeline A is selected to advance as a candidate solution.}

\smallskip \textbf{Step \#3. Crossover.} This step involves combining the genetic material of the two parent individuals to generate offspring. \textsc{FATE} will implement a \textit{single-point crossover} approach with probability $\alpha$ that involves splitting individuals into two parts and exchanging one of these with the other.
\steResearchQuestionBox{\textbf{Running Example \#3.} Crossover combines parts of selected pipelines to create new ones. For example, from pipeline A (Standard Scaling, Oversampling) and pipeline B (MinMax Scaling, Clustering), it produces offspring 1 (Standard Scaling, Clustering) and offspring 2 (MinMax Scaling, Oversampling). }

\smallskip \textbf{Step \#4. Mutation.} Mutation introduces random changes to an individual's genes, altering one or more key fairness-aware practices. \textsc{FATE} will mutate one of the offspring's practices with probability $\beta$ by randomly replacing it.

\steResearchQuestionBox{\textbf{Running Example \#4.}
To introduce diversity, mutation randomly alters a practice in an offspring with a probability $\beta$. For instance, offspring 1 mutates Clustering into Resampling. Afterward, with the offspring becoming the new population, the algorithm starts over by evaluating the fitness value of its individuals as in Step \#3. This iterative process of evaluation, selection, crossover, and mutation continues for multiple generations. Over time, the population of pipelines converges toward a near-optimal solution. For example, after 10 generations, \textsc{FATE} might identify a pipeline combining MinMax Scaling, Oversampling, and Matching, achieving a PS of 0.92 and an FS of 0.05. This result would balance predictive performance and fairness, demonstrating the effectiveness of the selected practices.}

\smallskip \textbf{Output.} \textsc{FATE} returns the near-optimal pipeline resulting from the evolutionary process of the genetic algorithm. Furthermore, it provides users with all the metrics computed for the evaluation of the fitness function, allowing them to understand why the selected pipeline should be applied. 
\section{Empirical Evaluation}

\begin{figure}
\centering
\includegraphics[width=.80\linewidth]{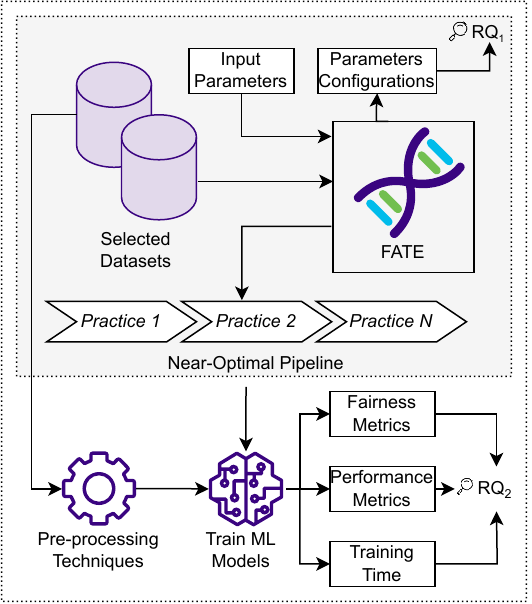}
\caption{Research Method.}
\label{fig:method}
\end{figure}

The \emph{goal} of the registered report is to investigate how the selection of fairness-aware practices during the Data Preparation phase of ML development affects the balance between fairness and performance. Specifically, the study will analyze how these practices influence the trade-offs that often arise when trying to optimize both fairness and model performance simultaneously, with the \emph{purpose} of understanding the extent to which these practices are able to support ML developers in mitigating bias. The study is viewed from the \emph{perspective} of both researchers and practitioners: researchers aim to advance the theoretical understanding of how fairness-aware practices can be systematically optimized in the \emph{`Data Preparation'} stage, while practitioners are interested in practical applications that ensure fairness without compromising model performance.

\subsection{Research Questions}
Our empirical study is organized around two main research questions (\textbf{RQ}s). The first \textbf{RQ} focuses on assessing the \emph{efficacy} of our proposed solution, delving into the fairness-accuracy trade-off that near-optimal pipelines may retain. We will conduct experiments to observe how \textsc{FATE} will perform under varying conditions, such as population sizes and numbers of generations, and evaluate how these parameters influence the fairness and accuracy outcomes of ML models trained with the optimized \emph{`Data Preparation'} pipelines.

\sterqbox{RQ\textsubscript{1}. Efficacy.}{To what extent can \textsc{FATE} select near-optimal configurations of fairness-aware practices?}

The second \textbf{RQ} aims to \emph{compare} our proposed solution with existing bias mitigation techniques, following the design of previous studies in the field \cite{peng2023fairmask}. Since \textsc{FATE} is designed to select the best fairness-aware \emph{`Data Preparation'} practices, we will compare the best configurations obtained from \textbf{RQ\textsubscript{1}} against established \emph{pre-processing bias mitigation methods \cite{pessach2022review, hort2024bias}}, evaluating both performance and fairness metrics.

\sterqbox{RQ\textsubscript{2}. Comparison.}{How does \textsc{FATE} compare to existing pre-processing techniques in mitigating bias?}

By investigating the efficacy of \textsc{FATE} in selecting fairness-aware \emph{`Data Preparation'} practices that optimize both fairness and accuracy and through careful comparison with state-of-the-art pre-processing bias mitigation techniques, our study aims to understand the role of fairness-aware practices in the fairness-accuracy trade-off against commonly-used bias mitigation methods.
Figure \ref{fig:method} shows the research method we plan to apply. The subsequent sections detail the experimental setup and the specific methods we will use to address our RQs. In terms of reporting, we will adhere to the \textsl{ACM/SIGSOFT Empirical Standards}.\footnote{Available at: \url{https://github.com/acmsigsoft/EmpiricalStandards}.} Based on the nature of our study, we will follow the \textsl{``General Standard''} guidelines.

\subsection{Experimental Setup}
This section explains the practices, datasets, models, and pre-processing techniques we will use to answer our \textbf{RQ}s.

\smallskip
\subsubsection{Practices Selection}
\textsc{FATE} will integrate with established fairness-aware \emph{`Data Preparation'} techniques \cite{voria2024mapping} to optimize both fairness and model performance. Below, we summarize the selected techniques for our experiments in \textbf{RQ\textsubscript{1}} and \textbf{RQ\textsubscript{2}}. This solution is adaptable to various combinations of fairness-focused \emph{`Data Preparation'} methods, specifically targeting improvements related to \textit{protected attributes}.

\begin{itemize}

\item \textit{Standard Scaling.} This technique \cite{Biswas2021981, Ilyas2022} normalizes numerical features to have a mean of zero and a standard deviation of one, preventing any single feature from disproportionately influencing the model.

\item \textit{MinMax Scaling.} By compressing numerical features to a set range (typically 0 to 1), MinMax scaling \cite{Biswas2021981, Ilyas2022} ensures features contribute equally to the model.

\item \textit{Resampling.} This method balances class representation through oversampling minority classes and undersampling majority classes, reducing bias toward overrepresented groups \cite{Iosifidis20191375, GonzalezZelaya20192086}. Our experiments use oversampling, undersampling, and stratified sampling variations.

\item \textit{Clustering.} Techniques like KMeans \cite{Gullo2022302} group data and enhance demographic representation by managing subgroups more effectively within the dataset.

\item \textit{Inverse Probability Weighting (IPW).} IPW assigns weights inversely proportional to class probabilities, mitigating selection bias and promoting proportional group representation \cite{pessach2022review, ipw}.

\item \textit{Matching.} By reordering data to create comparable groups, matching reduces biases, fostering fairer model predictions and group comparisons \cite{Biswas2021981}. \end{itemize}

\subsubsection{Datasets Selection}
To address \textbf{RQ\textsubscript{1}} and \textbf{RQ\textsubscript{2}}, we will use datasets containing sensitive attributes pertinent to fairness analysis. Building on the work of Fabris et al.~\cite{Fabris_2022}, which documents datasets widely used for fairness research, we have selected datasets from diverse domains, each with different sensitive attributes. This is essential to mitigate biased results and improve the robustness of our findings. 

The datasets we will select for this study are widely referenced in fairness literature \cite{chen2024fairness, de4966447examining} and span diverse application domains. Specifically, we will focus on three well-known datasets. The \textsl{German Credit} dataset \cite{german} contains data from approximately 1,000 loan applicants, with sensitive attributes such as \textsl{age'} and \textsl{sex’}. The \textsl{Heart Disease} dataset \cite{heart} includes around 1,000 patient records used to predict coronary artery disease, where \textsl{age'} and \textsl{sex’} are key attributes. Lastly, the \textsl{Adult} dataset \cite{adult} is designed to predict whether an individual’s annual income exceeds \$50,000, based on demographic and socioeconomic data, including protected attributes such as \textsl{race'} and \textsl{sex’}.

\subsubsection{Models Selection} 
To address our research questions, we plan to ML models to experiment with. To guide our selection, we draw upon prior research in the field \cite{Biswas2021981,udeshi2018automated, de4966447examining, chen2024fairness}. Specifically, we base our approach on the work of Hort et al. \cite{hort2024classifiersurvey}, who conducted a comprehensive review of the literature on ML fairness and identified the algorithms most commonly used in bias mitigation studies. We plan to select four algorithms: \textsl{Logistic Regression} (LR), \textsl{Linear Support Vector Classification} (SVC), \textsl{Random Forest} (RF), and \textsl{XGBoostClassifier} (XGB). Furthermore, these algorithms will be the ones used during the execution and evaluation of \textsc{FATE}. Particularly, we will run our solution separately for each model to compare results and answer our \textbf{RQs}.


\subsubsection{Pre-processing Techniques Selection}
Following the main objective of this study, in \textbf{RQ\textsubscript{2}} we will compare \textsc{FATE} to commonly used pre-processing bias mitigation techniques, following prior studies \cite{peng2023fairmask}. This comparison aims to identify which approach yields optimal results in terms of both accuracy and fairness. We plan to select three established pre-processing techniques aimed at mitigating bias. The first method we intend to verify is \textit{FairSMOTE} \cite{chakraborty2021bias}, a variation of the SMOTE algorithm \cite{chawla2002smote}, that is designed to generate synthetic samples for underrepresented classes. This method addresses class imbalances and improves the representation of minority classes in the dataset. Next, we will include \textit{Reweighing} \cite{kamiran2012data}, which rebalances data distributions by adjusting sample weights based on group characteristics, reducing disparities before model training. Lastly, we plan to apply the \textit{Disparate Impact Remover} \cite{feldman2015certifying} method, which modifies feature values to increase fairness while preserving the rank-ordering of values within each group, maintaining consistency across groups.

\subsection{Data Collection and Analysis}
This section describes the methods we plan to use to answer our two research questions. The previously described setup will be used for all the experiments conducted.

\smallskip
\subsubsection{\textbf{RQ\textsubscript{1}}: Evaluation of \textsc{FATE}'s Efficacy}
The first research question investigates how genetic algorithm parameters affect the selection of a near-optimal fairness-aware \emph{`Data Preparation'} pipeline. To address this, we will conduct a series of experiments that systematically vary \textsc{FATE}'s parameters \cite{whitley1994genetic} and record the specific results of each trial. Specifically, following the design of previous studies \cite{gonzalez2023preprocessing}, we will experiment with four key parameters: (1) \textit{population size}, (2) \textit{number of generations}, (3) \textit{crossover rate}, and (4) \textit{mutation rate}.

The genetic algorithm will be run for each combination of these parameters, and various metrics will be collected in each instance. Parameters include population sizes of 25, 50, 100, 250, and 500. The number of generations will be also set at 25, 50, 100, 250, and 500. Crossover rates and mutation rates will be set at 0\%, 25\%, 50\%, 75\%, and 100\%. We will gather data on \textit{fitness score}, \textit{performance score}, and \textit{fairness score} for each near-optimal solution generated during execution. We will compute two \textit{baselines} and collect the same metrics to compare our results. First, for each dataset selected, we will train the selected ML model without any \emph{`Data Preparation'} practice.  Secondly, we will run the same experiment with the \emph{`Data Preparation'} practices selected for the experimental setup. 
In both cases, to avoid possible bias and variability in performance evaluations, we will use K-Fold cross-validation to train and test ML models.

The comparison of the near-optimal solutions identified and the two baselines will provide insights into \textsc{FATE}'s behavior and how specific parameters influence each collected metric, enabling us to address \textbf{RQ\textsubscript{1}} understanding if our solution correctly selects near-optimal pipelines and by identifying the best parameter configuration for each measure.

\smallskip
\subsubsection{\textbf{RQ\textsubscript{2}}: Comparison of \textsc{FATE} with Pre-processing Techniques}

After identifying the best configurations in \textbf{RQ\textsubscript{1}}, we will perform an empirical evaluation comparing near-optimal sets of fairness-aware practices with existing pre-processing bias mitigation techniques. Specifically, we plan to apply these methods and \textsc{FATE}, using the best configuration, to the selected datasets within the experimental setup. Following this application, we will train the selected ML models for such evaluations by employing the K-Fold cross-validation strategy. Hence, we will execute our algorithm four times, one for each ML model selected, and collect data separately. We will gather three metrics: (1) \textit{execution time}, measured in seconds required to apply the \emph{`Data Preparation'} pipeline or pre-processing technique and train the model; (2) \textit{fairness} level of the trained model, assessed using metrics in Table \ref{tab:metrics}; and (3) \textit{predictive performance} of the trained model using metrics in the same table (\ref{tab:metrics}).

We will use statistical tests to evaluate whether \textsc{FATE}'s fairness and accuracy gains are significant compared to pre-processing techniques. These tests will validate our working hypothesis. Specifically, we intend to use a non-parametric test, such as the Wilcoxon Rank Sum Test \cite{conover1999practical}, with $\alpha = 0.05$, on the three metrics mentioned earlier. For this purpose, we formulate the following null hypotheses:

\textbf{H1a.}  \textit{There is no significant difference in terms of fairness between \textsc{FATE} and FairSMOTE.}

\textbf{H1b.}  \textit{There is no significant difference in terms of fairness between \textsc{FATE} and Reweighing.}

\textbf{H1c.}  \textit{There is no significant difference in terms of fairness between \textsc{FATE} and Disparate Impact Remover.}

\textbf{H2a.}  \textit{There is no significant difference in terms of performance between \textsc{FATE} and FairSMOTE.}

\textbf{H2b.}  \textit{There is no significant difference in terms of performance between \textsc{FATE} and Reweighing.}

\textbf{H2c.}  \textit{There is no significant difference in terms of performance between \textsc{FATE} and Disparate Impact Remover.}

\textbf{H3a.}  \textit{There is no significant difference in terms of execution time between \textsc{FATE} and FairSMOTE.}

\textbf{H3b.}  \textit{There is no significant difference in terms of execution time between \textsc{FATE} and Reweighing.}

\textbf{H3c.}  \textit{There is no significant difference in terms of execution time between \textsc{FATE} and Disparate Impact Remover.}

From a statistical perspective, we must account for the fact that if one of the null hypotheses is rejected, it implies that either FATE or the other technique is statistically superior to the other. Therefore, we will define a set of \textit{alternative hypotheses} that posit a statistically significant difference between our GA and other techniques across each measure.

We will reject the null hypothesis if $p < 0.05$. Alongside the non-parametric test, we employ the Vargha-Delaney test \cite{vargha2000critique} to quantify the effect size in the distributions of the selected metrics. This selection is justified by its advantages over mean-based tests, as it does not assume normality and is more robust to outliers and non-normal distributions. Additionally, it is applicable to ordinal data and offers a clear measure of effect size, statistically quantifying the tendency of one group to dominate the other. This provides more informative insights compared to analyzing distributions \cite{vargha2000critique}. Guided by the direction indicated by this test, we can interpret the alternative hypotheses in practical terms, determining whether our proposed genetic algorithm can surpass state-of-the-art pre-processing bias mitigation techniques, thereby addressing our \textbf{RQ\textsubscript{2}} and reaching our goal.

\section{Threats To Validity} This section addresses potential threats to the validity that could affect our empirical study and the mitigation strategies that we will implement through our research methods.

\textbf{Internal validity} focuses on ensuring that results genuinely reflect the factors we study. A key threat is the configuration of \textsc{FATE}'s parameters, particularly relevant to \textbf{RQ\textsubscript{1}}. To mitigate this, we will establish these parameters through preliminary experiments. Nonetheless, alternative configurations could lead to variations in results, affecting both performance and fairness outcomes. Additionally, GAs introduce randomness through initialization and probabilistic operations like selection, crossover, and mutation, which can also create variability. To address this, we will conduct multiple experiments, following methods from related studies \cite{gonzalez2023preprocessing}.

\textbf{External validity} concerns the generalizability of findings beyond our setup. \textsc{FATE} will be tested on a limited set of datasets, which may limit applicability across broader domains. We plan to address this by selecting datasets from diverse domains and tasks \cite{Fabris_2022}, but we acknowledge that further studies are needed to confirm the transferability of our results. To support replication and additional research, we will make all data and scripts publicly accessible.

\textbf{Construct validity} reflects how well the study’s measurements align with the constructs being assessed. One potential threat here is the choice of datasets used for model training and evaluation following the application of near-optimal practices. To counter this, we select widely used datasets \cite{Fabris_2022} that are relevant to our focus on fairness-performance trade-offs \cite{chen2024fairness,chakraborty2021bias,majumder2023fair}. Another consideration is our choice of fairness metrics (SPD, EOD, DI), which, while not exhaustive, are well-supported by current literature \cite{majumder2023fair, chen2024fairness} as robust measures for fairness evaluation. Similarly, the choice of PR AUC to evaluate predictive performance may influence the results. However, this metric has been shown to effectively complement commonly used metrics---precision, recall, accuracy, F1 score, ROC AUC---while being more efficient in handling classification tasks with unbalanced datasets, such as the ones used in the study. The selected ML models may also influence results; however, we will use well-established models employed in fairness research \cite{chen2024fairness}.

\textbf{Conclusion Validity} relates to the reliability of our conclusions. A primary threat involves the statistical tests planned \textbf{RQ\textsubscript{2}}, specifically the Wilcoxon \cite{conover1999practical} and Vargha-Delaney tests \cite{vargha2000critique}. These tests assume certain data distribution characteristics, and violating these assumptions could compromise the results. To address this, we will first assess the distribution of the project data to determine normality, ensuring that we select the most appropriate test for reliable conclusions.
\section{Conclusion}

This report evaluates how near-optimal fairness-aware practices can enhance both fairness and performance in ML models, potentially outperforming pre-processing bias mitigation techniques. Our contributions will be: (1) an optimization technique, \textsc{FATE}, to select \emph{`Data Preparation'} pipelines that balance both fairness and performance, (2) an analysis of how \textsc{FATE} manages the fairness-performance trade-off, and (3) an empirical study on the effects of \textsc{FATE}-selected pipelines with commonly used pre-processing bias mitigation techniques.

\section*{Acknowledgment}
We acknowledge the use of ChatGPT-4 to ensure linguistic accuracy and enhance the readability of this article. This work has been partially supported by the European Union -
NextGenerationEU through the Italian Ministry of University and
Research, Project PRIN 2022 PNRR "FRINGE: context-aware FaiR-
ness engineerING in complex software systEms" (grant n. P2022553SL,
CUP: D53D23017340001).

\balance
\bibliographystyle{IEEEtran}
\bibliography{references}

\end{document}